\documentclass[aps,prl,preprintnumbers,twocolumn,groupedaddress,showpacs]{revtex4}

\usepackage{latexsym}
\usepackage{amsmath}
\usepackage{amsfonts}
\usepackage{graphicx}
\usepackage{axodraw}

\newcommand{\bq}{\begin{eqnarray}}
\newcommand{\eq}{\end{eqnarray}}
\renewcommand{\l}{\langle}
\renewcommand{\r}{\rangle}

\begin{document}

\preprint{MZ-TH/11-41}
\title{\boldmath{NLO results for five, six and seven jets in electron-positron annihilation}}

\author{Sebastian Becker, Daniel G\"otz, Christian Reuschle, Christopher Schwan and Stefan Weinzierl}
\affiliation{Institut f{\"u}r Physik, Universit\"at Mainz, D-55099 Mainz, Germany}

\date{\today}

\begin{abstract}
We present next-to-leading order corrections in the leading colour approximation for
jet rates in electron-positron annihilation up to seven jets.
The results for the two-, three-, and four-jet rates agree with known results.
The NLO jet rates have been known previously only up to five jets.
The results for the six- and seven-jet rate are new.
The results are obtained by a new and efficient method based on subtraction and numerical integration.
\end{abstract}


\maketitle

%-----------------------------------------------------------
\section{Introduction}

Precise calculations for multi-parton final states are extremely important
for an accurate description of the physics at the LHC experiments.
The precision and accuracy is reached by including next-to-leading order (NLO)
QCD corrections in the theoretical prediction.
However, multi-leg NLO calculations are far from trivial.
In particular the virtual corrections for processes 
with many external legs have been considered to be a challenge for a long time.
In the past years there has been tremendous progress both in the refinement of the traditional Feynman diagram approach
and in the development of the unitarity method.
With the currently available
techniques one is able to obtain results for processes involving six- or seven-point one-loop functions.
Examples of such calculations are the NLO corrections to the processes
%$pp \rightarrow V + 3,4 \; \mbox{jets}$ \cite{Berger:2009zg,Berger:2009ep,Berger:2010vm,Berger:2010zx,Ita:2011wn,Ellis:2009zw,KeithEllis:2009bu}, 
$pp \rightarrow V + 3,4 \; \mbox{jets}$ \cite{Berger:2009zg,Ita:2011wn,Ellis:2009zw}, 
$pp \rightarrow WW + 2 \; \mbox{jets}$ \cite{Melia:2010bm},
$pp \rightarrow t\bar{t} + 2 \, \mbox{jets}$ \cite{Bevilacqua:2010ve},
$pp \rightarrow t\bar{t} b \bar{b}$ \cite{Bevilacqua:2009zn,Bredenstein:2009aj} or
$e^+ e^- \rightarrow 5 \; \mbox{jets}$ \cite{Frederix:2010ne}.
Most of these calculations are based on the unitarity method.
The results involving one-loop seven-point functions have been obtained only in the leading-colour approximation.

In this letter we report on results obtained by a completely different method.
We calculate the NLO corrections by numerical 
%methods \cite{Becker:2010ng,Assadsolimani:2010ka,Assadsolimani:2009cz,Gong:2008ww,Anastasiou:2007qb,Nagy:2006xy,Nagy:2003qn,Soper:2001hu,Soper:1999xk,Soper:1998ye}. 
methods \cite{Becker:2010ng,Assadsolimani:2010ka,Assadsolimani:2009cz,Gong:2008ww,Anastasiou:2007qb,Nagy:2006xy,Soper:2001hu}. 
To this aim we extend the subtraction method from the real emission part to the virtual part. 
This renders the integration over the loop momentum finite
and we perform the loop integration together with the phase space integration in one Monte Carlo integration.
In this letter we demonstrate the potential of this method.
We calculate the NLO corrections for the process $e^+e^- \rightarrow n \; \mbox{jets}$ in the leading colour 
approximation up to $n=7$.

The LEP experiments have collected data on the jet rates in electron-positron annihilation. This data together
with a theoretical calculation can be used to extract the strong coupling $\alpha_s$. 
The precise value of $\alpha_s$ is an important input parameter for all LHC experiments.

The results for the two-, three-, and four-jet rates have been known for a long time and we use them to verify
the validity of our method.
The NLO corrections to the five-jet rate have been calculated recently.
The results on the six- and seven-jet rate are new. 
Note that the NLO corrections to the seven-jet rate require the calculation of a one-loop eight-point function.
To the best of our knowledge this is the first time that physical observables 
depending on a one-loop eight-point function have been calculated.

We have written a numerical programme based on recursion relations. 
This programme can in principle calculate the NLO corrections to $n$-jet observables in the leading colour approximation for all $n$.
Changing the number $n$ amounts to changing a single line in the numerical programme.
Within our method we have a very moderate growth in the CPU time per evaluation as $n$ increases.
The practical limitations arise from the fact that the number of evaluations required to reach a certain accuracy increases with $n$.
This behaviour is already present at the Born level and not inherent to our method.

The application of our method is not limited to electron-positron annihilation. 
The calculation of processes in hadron-hadron collisions requires only minor modifications and will be addressed in the near future.

% ------------------------------------------------------------
\section{The numerical method}

In this section we briefly outline the method for the numerical computation of NLO corrections.
More details can be found in \cite{Becker:2010ng}.

In electron-positron annihilation 
the contributions at leading and next-to-leading order for an infrared-safe $n$-jet observable $O$ are given as
\bq
\l O \r^{\mathrm{LO}} & = & 
 \int\limits_n O_n d\sigma^{\mathrm{B}},
 \\
\l O \r^{\mathrm{NLO}} & = & 
 \int\limits_{n+1} O_{n+1} d\sigma^{\mathrm{R}} + \int\limits_n O_n d\sigma^{\mathrm{V}}.
 \nonumber
\eq
Here a rather condensed notation is used. $d\sigma^{\mathrm{B}}$ denotes the Born
contribution,
whose matrix elements are given by the square of the Born amplitudes with $(n+2)$ particles.
Similar, $d\sigma^{\mathrm{R}}$ denotes the real emission contribution,
whose matrix elements are given by the square of the Born amplitudes with $(n+3)$ particles.
$d\sigma^{\mathrm{V}}$ gives the virtual contribution, whose matrix elements are given by the interference term
of the one-loop amplitude with $(n+2)$ particles, with the corresponding
Born amplitude.
The individual contributions at next-to-leading order 
are divergent and only their sum is finite.
We rewrite the NLO contribution as
\bq
\l O \r^{\mathrm{NLO}} & = & 
 \l O \r^{\mathrm{NLO}}_{\mathrm{real}} 
 + \l O \r^{\mathrm{NLO}}_{\mathrm{virtual}} 
 + \l O \r^{\mathrm{NLO}}_{\mathrm{insertion}}.
\eq
In order to render the real emission contribution finite, such that the phase space integration
can be performed by Monte Carlo methods, one subtracts a suitably chosen piece
\cite{Catani:1997vz,Dittmaier:1999mb,Phaf:2001gc,Catani:2002hc}:
\bq
\l O \r^{\mathrm{NLO}}_{\mathrm{real}} & = & 
 \int\limits_{n+1} \left( O_{n+1} d\sigma^{\mathrm{R}} - O_n d\sigma^{\mathrm{A}} \right).
\eq
This is the standard application of the subtraction method.
Within our method we use the same idea for the virtual part:
\bq
\label{virtual_part}
\lefteqn{
\l O \r^{\mathrm{NLO}}_{\mathrm{virtual}} =  
 2 \int d\phi_n 
 \;\mbox{Re}\; \int \frac{d^4k}{(2\pi)^4}
} & & \nonumber \\
 & &
 \left[\left.{\cal A}^{(0)}\right.^\ast 
 \left( {\cal G}_{\mathrm{bare}}^{(1)} - {\cal G}_{\mathrm{soft}}^{(1)} 
        - {\cal G}_{\mathrm{coll}}^{(1)} - {\cal G}_{\mathrm{UV}}^{(1)} \right) \right] O_{n}.
\eq
Here, ${\cal G}^{(1)}_{\mathrm{bare}}$ denotes the integrand of the bare one-loop amplitude:
\bq
{\cal A}^{(1)}_{\mathrm{bare}} & = & \int \frac{d^4k}{(2\pi)^4} {\cal G}^{(1)}_{\mathrm{bare}}.
\eq
${\cal G}_{\mathrm{soft}}^{(1)}$, ${\cal G}_{\mathrm{coll}}^{(1)}$ and ${\cal G}_{\mathrm{UV}}^{(1)}$
are local subtraction terms, approximating the soft, collinear and ultraviolet singularities of the integrand.
The subtraction terms have to be added back, which is done with the term
\bq
\label{insertion_part}
\l O \r^{\mathrm{NLO}}_{\mathrm{insertion}} & = & 
 \int\limits_{n} O_{n} \left( {\bf I} + {\bf L} \right) \otimes d\sigma^B.
\eq
The insertion operator ${\bf I}$ contains the integrated subtraction terms from the real emission part,
the insertion operator ${\bf L}$ contains the integrated subtraction terms from the virtual part plus the ultraviolet
counterterm from renormalisation.
By construction, all three contributions $\l O \r^{\mathrm{NLO}}_{\mathrm{real}}$, $\l O \r^{\mathrm{NLO}}_{\mathrm{virtual}}$
and $\l O \r^{\mathrm{NLO}}_{\mathrm{insertion}}$ are finite.
It should be noted that all quantities are defined at the amplitude level and are calculated using recurrence relations.
This allows for a fast implementation.

Let us discuss eq.~(\ref{virtual_part}) in more detail.
The subtraction terms ensure that the result of the loop integration is finite.
This does not yet imply that we can perform the loop integration numerically by integrating the four loop momentum components from
minus infinity to plus infinity.
The integrand still has singularities for certain values of the loop momentum, i.e. if one or more of the propagators in the loop goes on-shell.
This can be avoided by deforming the loop integration into the complex space,
which is the second essential ingredient of our method.
We devised and implemented an algorithm for the contour deformation in two variants.
One variant is based on an auxiliary Feynman parametrisation \cite{Nagy:2006xy,Becker:2010ng}, the other one
works directly in loop momentum space \cite{Gong:2008ww}.
The two variants yield identical results, with the direct loop momentum deformation method being more efficient for higher jet rates.

% ------------------------------------------------------------
\section{Jet rates in electron-positron annihilation}

In this paper we consider jet rates in electron-positron annihilation.
The jets are defined by the Durham jet algorithm \cite{Stirling:1991ds}.
For the Durham jet algorithm the
resolution variable is defined by
\bq
 y_{ij} & = & \frac{2 \min(E_i^2, E_j^2) \left( 1 - \cos \theta_{ij} \right)}{Q^2},
\eq
where $E_i$ and $E_j$ denote the energies of particles $i$ and $j$, and $\theta_{ij}$ is the angle
between the three-momenta $\vec{p}_i$ and $\vec{p}_j$.
$Q$ is the centre-of-mass energy.

The production rate for $n$-jet events in electron-positron annihilation is given as
the ratio of the cross section for $n$-jet events divided by the total hadronic cross section
\bq
 R_n(\mu) & = &  \frac{\sigma_{n-\mathrm{jet}}(\mu)}{\sigma_{\mathrm{tot}}(\mu)}.
\eq
The arbitrary renormalisation scale is denoted by $\mu$.
The jet rates can be calculated within perturbation theory.
The perturbative expansion of the jet rate reads
\bq
\lefteqn{
 R_n(\mu) = 
} & &   
 \\
& &
 \left( \frac{\alpha_s(\mu)}{2\pi} \right)^{n-2} \bar{A}_n(\mu)
    + \left( \frac{\alpha_s(\mu)}{2\pi} \right)^{n-1} \bar{B}_n(\mu)
    + {\cal O}(\alpha_s^n).
 \nonumber
\eq
In practise the numerical programme computes the quantities
\bq
\lefteqn{
 \frac{\sigma_{n-\mathrm{jet}}(\mu)}{\sigma_{0}(\mu)} = 
} & &
 \\
 & &
    \left( \frac{\alpha_s(\mu)}{2\pi} \right)^{n-2} A_n(\mu)
    + \left( \frac{\alpha_s(\mu)}{2\pi} \right)^{n-1} B_n(\mu)
    + {\cal O}(\alpha_s^n), 
 \nonumber
\eq
normalised to $\sigma_0$, which is the LO cross section for $e^+ e^- \rightarrow \mbox{hadrons}$,
instead of the normalisation to $\sigma_{\mathrm{tot}}$.
From the expansion 
of the total hadronic cross section $\sigma_{\mathrm{tot}}$
\bq
 \sigma_{\mathrm{tot}}(\mu) & = & 
 \sigma_0(\mu) \left( 1 + \frac{3}{2} C_F \frac{\alpha_s(\mu)}{2\pi} 
                   + {\cal O}(\alpha_s^2) \right),
\eq
we obtain the relations
between the coefficients $A_n$, $B_n$ and
the coefficients $\bar{A}_n$, $\bar{B}_n$:
\bq
 \bar{A}_n = A_n,
 & & 
 \bar{B}_n = B_n - \frac{3}{2} C_F A_n.
\eq
It is sufficient to calculate the coefficients $\bar{A}_{n}$ and $\bar{B}_{n}$
for a fixed renormalisation scale $\mu_0$, which can be taken conveniently to be equal to the
centre-of-mass energy $\mu_0=Q$.
The scale variation can be restored from the renormalisation group equation
\bq
\label{RGE_alpha_s}
 \mu^2 \frac{d}{d\mu^2} \left( \frac{\alpha_S}{2\pi} \right)
 & = & 
 - \frac{1}{2} \beta_0 \left( \frac{\alpha_S}{2\pi} \right)^2
 + {\cal O}(\alpha_s^3),
 \nonumber \\
 \beta_0 & = & \frac{11}{3} C_A - \frac{4}{3} T_R N_f.
\eq
The values of the coefficients $\bar{A}_n$ and $\bar{B}_n$
at a scale $\mu$ are then obtained from the ones at the scale $\mu_0$ by
\bq
 \bar{A}_n(\mu) & = & \bar{A}_n(\mu_0),
 \nonumber \\
 \bar{B}_n(\mu) & = & \bar{B}_n(\mu_0) + \frac{(n-2)}{2} \beta_0 \ln\left(\frac{\mu^2}{\mu_0^2}\right) \bar{A}_n(\mu_0).
\eq

We can expand the perturbative coefficients $A_n$ and $B_n$ in $1/N_c$:
\bq
 A_n & = & N_c \left( \frac{N_c}{2} \right)^{n-2} \left[ A_{n,\mathrm{lc}} + {\cal O}\left( \frac{1}{N_c} \right) \right],
 \nonumber \\
 B_n & = & N_c \left( \frac{N_c}{2} \right)^{n-1} \left[ B_{n,\mathrm{lc}} + {\cal O}\left( \frac{1}{N_c} \right) \right].
\eq
In this article we calculate the coefficients $A_{n,\mathrm{lc}}$ and $B_{n,\mathrm{lc}}$ for $n \le 7$.

% ------------------------------------------------------------
\section{Numerical results}

In this section we present the numerical results. We calculate the leading order coefficient $A_{n,\mathrm{lc}}$ and the
next-to-leading order coefficient $B_{n,\mathrm{lc}}$ for $n \le 7$ for the fixed renormalisation scale
equal to the centre-of-mass energy $\mu_0=Q$.
We take the centre-of mass energy to be equal to the mass of the $Z$-boson $Q=m_Z$.
The calculation is done with five massless quark flavours.

\begin{figure}[t]
\begin{center}
\includegraphics[bb= 125 460 490 710,width=0.5\textwidth]{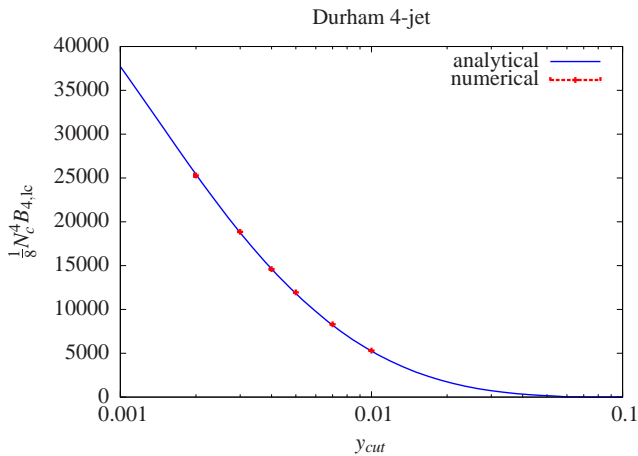}
\end{center}
\caption{
Comparison of the NLO corrections to the four-jet rate between the numerical calculation and an analytic calculation.
The error bars from the Monte Carlo integration are shown and are almost invisible.
}
\label{fig_4jet}
\end{figure}

We first compare our approach with well-known results for two, three and four jets \cite{Weinzierl:1999yf,Weinzierl:2010cw}.
\begin{figure}[t]
\begin{center}
\includegraphics[bb= 125 460 490 710,width=0.5\textwidth]{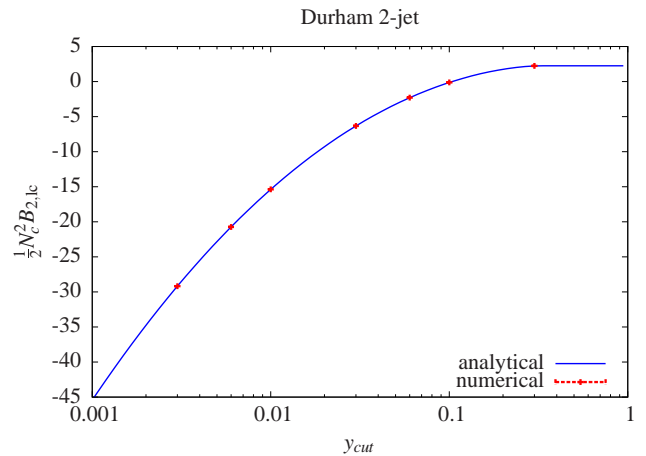}
\end{center}
\caption{
Comparison of the NLO corrections to the two-jet rate between the numerical calculation and an analytic calculation.
The error bars from the Monte Carlo integration are shown and are almost invisible.
}
\label{fig_2jet}
\end{figure}
Figure~(\ref{fig_2jet}) shows a comparison for the two-jet rate for the NLO coefficient $B_{2,\mathrm{lc}}$ between
the numerical calculation and an analytic calculation.
It should be noted that for the two-jet case the dependence on $y_{\mathrm{cut}}$ enters only through the real
emission contribution.
We observe an excellent agreement.
\begin{figure}[t]
\begin{center}
\includegraphics[bb= 125 460 490 710,width=0.5\textwidth]{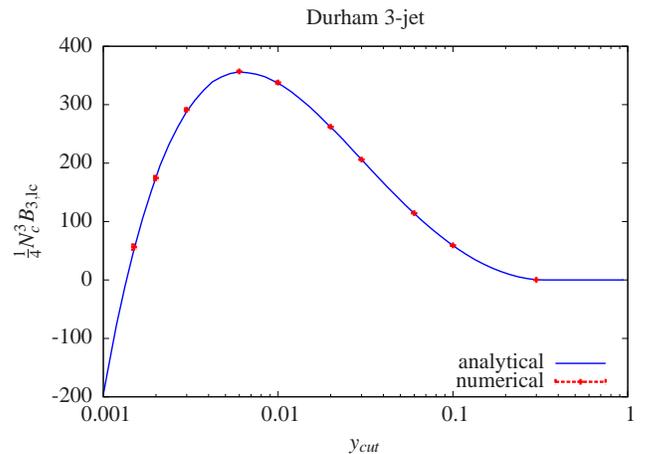}
\end{center}
\caption{
Comparison of the NLO corrections to the three-jet rate between the numerical calculation and an analytic calculation.
The error bars from the Monte Carlo integration are shown and are almost invisible.
}
\label{fig_3jet}
\end{figure}
The corresponding comparisons are shown for the three-jet rate in figure~(\ref{fig_3jet}) and for the four-jet rate in figure~(\ref{fig_4jet}).
Again we observe an excellent agreement.
\begin{table}
\begin{center}
\begin{tabular}{|l|c|c|}
\hline
 & & \\
 $y_{\mathrm{cut}}$ & $\frac{N_c^4}{8}  A_{5,\mathrm{lc}}$ & $\frac{N_c^5}{16} B_{5,\mathrm{lc}}$ \\
 & & \\
\hline 
 $0.002$ & $\left( 5.0529 \pm 0.0004 \right) \cdot 10^{3}$ & $ \left( 4.275 \pm 0.006 \right) \cdot 10^{5} $ \\
 $0.001$ & $\left( 1.3291 \pm 0.0001 \right) \cdot 10^{4}$ & $\left( 1.050 \pm 0.026 \right) \cdot 10^{6}$ \\
 $0.0006$ & $\left( 2.4764 \pm 0.0002 \right) \cdot 10^{4}$ & $\left( 1.84 \pm 0.15 \right) \cdot 10^{6}$ \\
\hline
\hline
 & & \\
 $y_{\mathrm{cut}}$ & $\frac{N_c^5}{16} A_{6,\mathrm{lc}}$ & $\frac{N_c^6}{32} B_{6,\mathrm{lc}}$ \\
 & & \\
\hline 
 $0.001$ & $( 1.1470 \pm 0.0002 ) \cdot 10^{5}$ & $( 1.46 \pm 0.04 ) \cdot 10^{7}$ \\
 $0.0006$ & $( 2.874 \pm 0.002 ) \cdot 10^{5}$ & $( 3.88 \pm 0.18 ) \cdot 10^{7}$ \\
\hline
\hline
 & & \\
 $y_{\mathrm{cut}}$ & $\frac{N_c^6}{32} A_{7,\mathrm{lc}}$ & $\frac{N_c^7}{64} B_{7,\mathrm{lc}}$ \\
 & & \\
\hline 
 $0.0006$ & $\left( 2.49 \pm 0.08 \right) \cdot 10^{6}$ & $( 5.4 \pm 0.3 ) \cdot 10^{8}$ \\
\hline
\end{tabular}
\caption{\label{table_567jet}
Perturbative coefficients for the five-jet rate, six-jet rate and seven-jet rate.
}
\end{center}
\end{table}
The results for the five-, six- and seven-jet rates are given in table~(\ref{table_567jet}) for a few selected values of the jet parameter
$y_{\mathrm{cut}}$. Note that a reasonably large value for the higher jet rates requires a small jet parameter $y_{\mathrm{cut}}$.

Let us have a closer look at the performance of our method.
\begin{figure}[t]
\begin{center}
\includegraphics[bb= 125 460 490 710,width=0.5\textwidth]{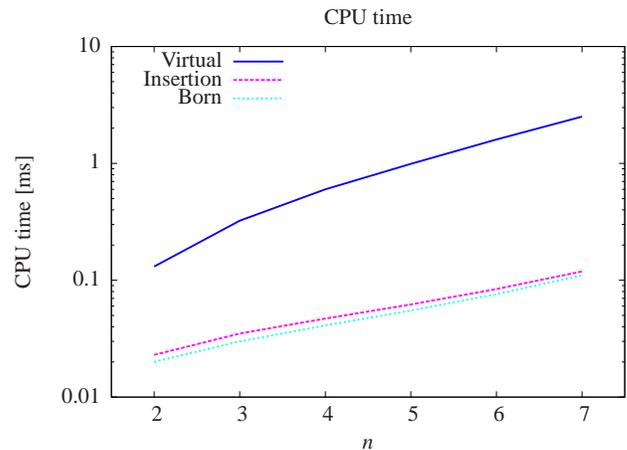}
\end{center}
\caption{
CPU time required for one evaluation of the Born contribution, 
the insertion term and the virtual term, respectively, as a function of the number of the final state partons $n$.
The times are taken on a single core of a standard PC.
}
\label{fig_cpu}
\end{figure}
Figure~(\ref{fig_cpu}) shows the CPU time required for one evaluation of the Born contribution, 
the insertion term in eq.~(\ref{insertion_part}) and the virtual term in eq.~(\ref{virtual_part}) as a function of the number of final state partons $n$.
One clearly sees that the insertion term is almost as cheap as the Born contribution.
But more important is the fact that for all contributions the CPU time per evaluation 
increases only very moderately as a function of $n$.
Within our method all three contributions should scale asymptotically as $n^4$ \cite{Kleiss:1988ne}. 
Fitting a polynomial function $n^\alpha$ 
to the numbers shows that this is indeed the case.
In principle we could further improve the $n^4$ scaling behaviour to $n^3$ by replacing the four-gluon vertex by two effective
three-valent vertices \cite{Duhr:2006iq}.
We emphasise that the virtual part has the same scaling behaviour as the Born contribution.
This moderate growth imposes almost no restrictions on the number of final state partons to which our method can be applied.
The practical limitations arise from the fact that the number of evaluations required to reach a certain accuracy increases with $n$.
This behaviour is already present at the Born level and not inherent to our method.
Altogether, the calculation of the seven-jet rate takes a few days on a cluster with 200 cores.

%-----------------------------------------------------------
\section{Conclusions}

In this letter we reported on NLO corrections in the leading colour approximation for jet rates in electron-positron annihilation up to seven jets.
The calculation is based on a new and powerful method, where the loop integration is performed numerically.
To the best of our knowledge this is the first time that physical observables depending on a one-loop eight-point function have been calculated.
We are planning to extend this method to hadron-hadron collisions and to include all terms in the colour expansion.

%-----------------------------------------------------------

%\bibliography{/home/stefanw/notes/biblio}
%\bibliographystyle{/home/stefanw/latex-style/h-physrev5}

\end{document}